\documentclass[aps,prc,twocolumn,showpacs,preprintnumbers,amsmath,amssymb]{revtex4}

\usepackage{graphicx}
\usepackage{dcolumn}
\usepackage{bm}
\usepackage{color}
\usepackage{float}

\begin{document}

\title{Evidence of a new shell closed nucleus governing slow quasi-fission }
\author{ A. Pal$^{1}$}
\email{asimpal@barc.gov.in}
\author{  S. Santra$^{1,2}$}
\author{ A. Kundu $^{1}$}
\author{ D. Chattopadhyay$^{1}$}
\author{ P.C. Rout $^{1,2}$}
\author{ Ramandeep Gandhi $^{1,2}$}
\author{ P. N. Patil $^{3}$}
\author{ R. Tripathi $^{2,4}$}
\author{ B. J. Roy $^{1,2}$}
\author{ Y. Sawant $^{1}$}
\author{ T.N. Nag $^{2,4}$}
\author{ Abhijit Baishya $^{1,2}$}
\author{ T. Santhosh $^{1,2}$}
\author{ P.K. Rath $^{5}$}
\author{ N. Deshmukh $^{6}$}

\address{$^1$Nuclear Physics Division, Bhabha Atomic Research Centre, Mumbai - 400085, India}
\address{$^2$Homi Bhabha National Institute, Anushaktinagar, Mumbai - 400094, India}
\address{$^3$Department of Physics, KLE Technological University,Hubballi-580031 , India}
\address{$^4$Radio Chemistry Division, Bhabha Atomic Research Centre, Mumbai - 400085, India}
\address{$^5$Centurion University of Technology and Management, Paralakhemundi, Odisha - 576104, India}
\address{$^6$School of Sciences, PP Savani University, Dhamdod, Kosamba, Surat  394 125, India}

\date{\today}

\begin{abstract}
Mass distributions of fission fragments arising from the slow quasi-fission process have been derived by comparing the measured distributions with the theoretical distributions based on compound nuclear fission model for several reactions. The mass-distributions corresponding to quasi-fission events for all the systems show the following common features: (1) they are double peaked with fixed peak-centroids and nearly same width at different incident energies, (2) the yield of quasi-fission events decreases with the increasing projectile energy, and (3) peak corresponding to lighter fragment is observed at A $\sim$ 96 for all the systems, whereas the peak of heavier fragment increases linearly with the mass of the di-nuclear system. All the above observations are quite similar to the ones observed in well known asymmetric fission of actinides, thus providing clear evidences of shell effect in slow quasi-fission where the lighter fragment is possibly nuclei around $^{96}$Zr, a new doubly magic nucleus. This finding has great implications in the study of nuclear reactions, structure and particularly in super-heavy element synthesis where quasi-fission is synonymous.
\end{abstract}
\maketitle

While synthesizing super-heavy elements (SHE) using fusion reactions \cite{og04,og06,og07,og10,og15,dmitriev05}, a pre-equilibrium fission reaction mechanism, generically named as Quasi-fission (QF) \cite{jp75,rb82,jt85} is known since mid 1970 as a cause for the suppression of SHE formation. Since then many aspects of QF have been explored experimentally by measuring fission fragment (FF) mass and angular distributions \cite{sk85,hinde95,rama90,hinde08,hinde18,rafiei08,prasad16,kozulin16,itkis15,mgitkis07,asen22,morjean08,thomas08,yadav12,tbanerjee20,banerjee11,ghosh05,ghosh09,ghosh041} and theoretically by developing many macroscopic and microscopic dynamical models\cite{torres01,aritomo12,sim12,simenel12,seki16}. From the above studies, one can broadly classify the QF process into two categories: Fast and Slow quasi-fission. The fast quasi-fission (FQF) which is generally observed  in reactions with heavy targets and projectiles, having charge product ($Z_pZ_t$) more than 1500, are characterized by very asymmetric mass -distributions, very fast time scales ($\sim 10^{-20}$ s) and presence of mass-angle correlation \cite{hinde08,hinde18,rafiei08,prasad16,kozulin16,itkis15,mgitkis07,asen22,morjean08}. In contrast, the slow quasi-fission (SQF) which is observed in reactions involving much lighter projectiles such as $^{9}$Be,$^{11}$B,$^{12}$C,
$^{16}$O etc. with actinide targets, is characterized mainly by a time scale intermediate to FQF and Compound nuclear fission (CNF), nearly symmetric mass distributions, absence of mass-angle correlations, larger mass width from the most symmetric entrance channel populating the same compound nucleus, sudden enhancement  in the mass-width at lower incident energies and most importantly larger angular anisotropy compared to the statistical model predictions \cite{thomas08,yadav12,tbanerjee20,banerjee11,ghosh05,ghosh09,ghosh041}.

\begin{figure}
\begin{center}
\includegraphics[scale=.67]{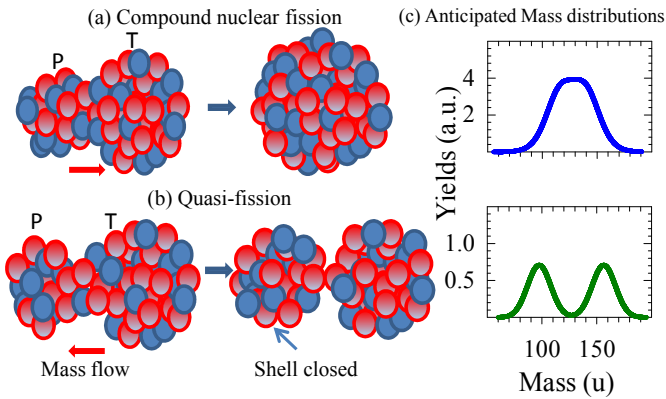}
\caption{\label{fig:fission} (Color online) Illustration of (a)compound nuclear fission process,(b) quasi-fission process and (c) expected mass distributions.}
\end{center}
\end{figure}

Recently the microscopic shell effect has been invoked to explain the mass distributions in quasi-fission process \cite{mitkis04,wakhle14,oberacker14,simenel21,morjean17}. Theoretically, based on a recent time dependent Hartree Fock calculation on $^{50}$Ca+$^{176}$ Yb reaction partners \cite{simenel21} forming $^{226}$Th composite system, it was found that the same deformed shell $Z_H \sim 54$ as that of S-II mode \cite{cb08,gc18,santra14,pal18,apal21} in asymmetric fission of actinides is responsible for stopping mass equilibration in the fast quasi-fission process, without allowing the system to form a compound nucleus. Experimentally also evidence for the role of proton shell closure in fast quasi-fission reactions have been observed recently\cite{morjean17}. However till date, there is no investigation on the role of shell effect in SQF reaction mechanism. Actually in a slow quasi-fission reaction process, the nucleon exchange starts happening from heavier (T) to lighter (P)colliding partner, in contrast to the compound nucleus formation process, as illustrated in Fig. \ref{fig:fission} (a-b). We can strongly anticipate that if one of the fragments becomes shell closed during the mass equilibration process in SQF process, the di-nuclear system breaks into two fragments resulting in a double humped mass distribution, while mass-distributions in compound nuclear fission process are superimposition of several fission modes (super-long, S-I,S-II etc.) depending on the excitation energy and angular momenta populated in the system (Fig. \ref{fig:fission}(c)). However experimentally it is impossible to distinguish the fragments arising from CNF and SQF processes unless we take help from theoretical models incorporating CNF process.

In the present letter, FF mass-distributions have been measured in $^{19}$F + $^{238}$U reaction and compared with the theoretical model calculations. From the differences observed between the measured data and the calculation, the mass distributions and the probabilities corresponding to QF process have been obtained at different projectile energies for the present system and the other systems for which FF mass distributions are available in literature. From the derived mass-distributions of the QF process for different systems, we attempt to search for the quantum shell effect in the slow quasi-fission process for the first time, towards the development of complete understanding of quasi-fission mechanism and to improve the models in reaching reliable predictive capacities of SHN formation cross-sections.


\begin{figure*}
\begin{center}
\includegraphics[scale=.7]{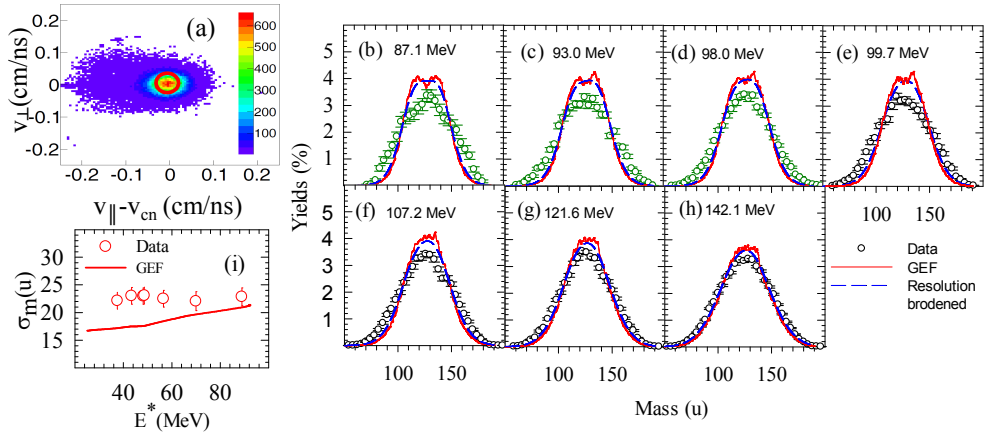}
\caption{\label{fig:mass} (Color online)(a) Typical `$v_\parallel$-$v_{cn}$' versus `$v_\perp$', obtained for $^{19}$F+$^{238}$U reaction at $E_{\rm beam}=142.1$ MeV. (b-d) Earlier and (e-h) presently measured FF mass distributions, where data points are shown by symbols, the red solid lines represent the mass distribution obtained from the model GEF and the blue dashed lines represent the distribution obtained after broadening the actual distribution by experimental mass resolution. (i) Width of the measured (circles) and calculated (solid line) mass distributions.}
\end{center}
\end{figure*}

\begin{figure}
\begin{center}
\includegraphics[scale=.8]{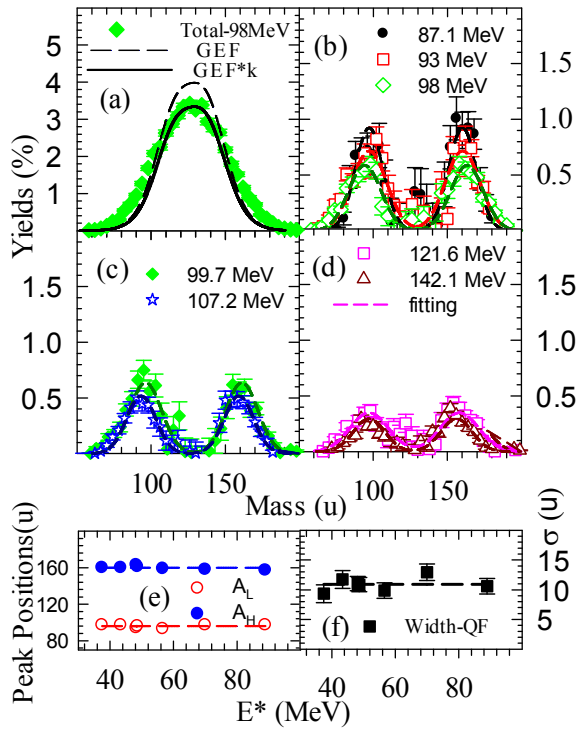}
\caption{\label{fig:mass-qf} (Color online) (a)Comparison of measured, actual and normalized calculated distributions, (b-d) Derived mass distributions corresponding to QF process at different projectile energies, (e-f) Peak positions and width of the QF modes.}
\end{center}
\end{figure}

Fission fragments have been measured for the $^{19}$F + $^{238}$U reaction using pulsed beam (FWHM $\sim$ 1 ns) of energies ranging from 98.7 to 142.1 MeV at 14-UD BARC-TIFR Pelletron-Linac facility, Mumbai. The details of the experimental setup and analysis procedure is same as in Ref\cite{apal21}.
 The full momentum transfer events have been selected using a tight gate on $v_\parallel-v_{cn}$ versus $v_\parallel$ plot as shown in Fig.~\ref{fig:mass}(a) and analyzed for the present work. From the measured velocities of the fragments in the centre of mass frame,  massess of the fragments have been derived on event by event basis. The whole mass distribution spectrum is normalized to 200 $\%$ and shown in Fig. \ref{fig:mass}(d-g) by black circles. Our previously measured data on the same reaction system at lower beam energies are also shown in the same Fig. \ref{fig:mass}(a-c) by green circles.
Now to understand the data, FF mass distributions are calculated using a semi-empirical model code GEF \cite{schmidt16,kh16} which calculates the distributions (normalized to 200 $\%$) for compound nuclear fission process. It could be observed that the measured distributions is little wider than the calculated ones shown by red solid line in Fig. \ref{fig:mass}, specially at lower beam energies. The calculated peak height overshoots the measured peak height for all the cases, more prominently for the beam energies up to 107.2 MeV. Simultaneously the tail part of the calculated distributions undershoots the measured data. To rule out the fact that the broadening of the measured data is due to the limited mass resolution of the experimental setup, the calculated distribution is broadened taking into account the mass resolution of the experimental setup, sigma$\sim$6 u, as shown by blue dashed lines in Fig. \ref{fig:mass}. One can notice that even after the inclusion of broadening the theoretical distributions fail to explain the much wider measured data. Thus the above analysis hints at possibilities of having admixture of quasi fission along with the compound nuclear fission. It may be mentioned that as there is no wing like structure in the asymmetric mass region of the experimental data and no mass-angle correlation is observed as per our expectations, the presence of fast quasi-fission is ruled out.
\begin{figure*}
\begin{center}
\includegraphics[scale=.9]{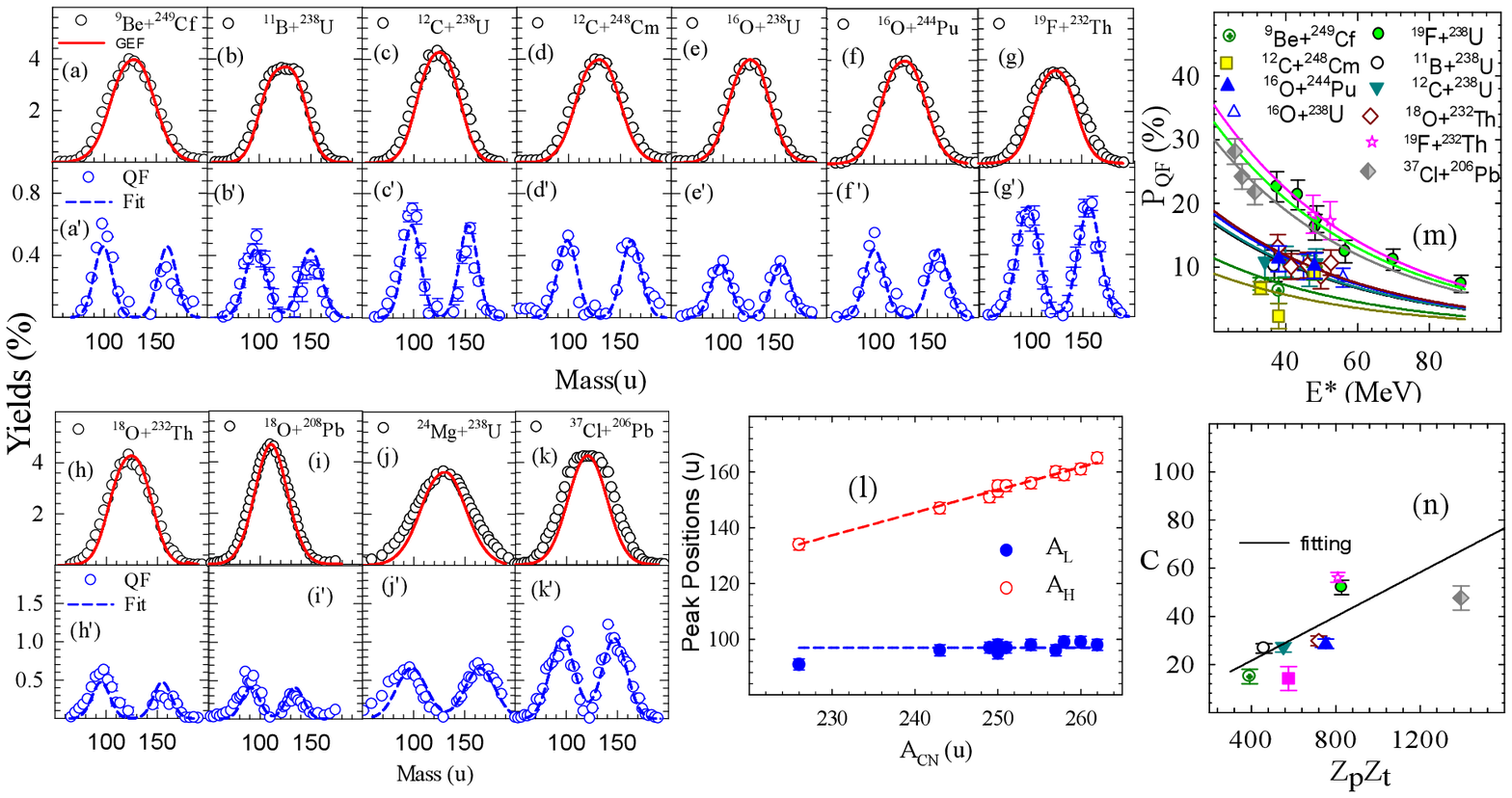}
\caption{\label{fig:qf-lit} (Color online) (a-k) Comparison of measured data and normalized GEF calculations for reaction systems available in literature, (a'-k') Derived mass distributions for QF mode for respective system, (l)Peak positions of light and heavy fragments of QF mass distributions, (m) Quasi-fission probability obtained as a function of excitation energy for different reaction systems. Fitting of the probabilities for each system is shown by solid lines with colors matched with the symbols. (n) The fitting constant-C for different systems. }
\end{center}
\end{figure*}


Now the width of the FF mass distribution as a function of compound nuclear excitation energy for the same reaction has been obtained and compared with the ones obtained from GEF calculations as shown in Fig \ref{fig:mass} (i). It can be observed that the deviation in width for the present system at lower excitation energies is much more than the width at higher energies, indicating the more contributions of slow quasi-fission at below barrier energies. As the beam energy is increased the contributions from quasi-fission becomes much smaller.


From the above facts and as already the evidence of slow quasi-fission in reactions with much lighter projectiles like $^{9}$Be, $^{12}$C and $^{16}$O were seen earlier \cite{tbanerjee20}, one can assure that the measured mass distributions contain contributions from both the processes: compound nuclear fission and slow quasi-fission, where the mass distributions following compound nuclear fission processes could be calculated reliably using theoretical models. As quasi-fission is expected to split in asymmetric mass divisions \cite{simenel21} and the symmetric split is expected to originate from compound nuclear fission, the measured distributions and the calculated distributions corresponding to compound nuclear fission are matched at fragment mass=0.5$\times$ compound nuclear mass, after multiplying the calculated distributions by a factor k, followed by subtracting it from the measured data resulting in the mass distributions only due to QF process. For illustration, the measured data (filled rhombus), actual calculated distribution (black dashed line) and normalized calculated distribution (black solid line) are shown in Fig. \ref{fig:mass-qf}(a) for 98 MeV projectile energy. The QF mass-distribution derived from the subtraction of data (filled rhombus)  and normalized calculation (black solid line) for the same 98 MeV projectile energy is shown by hollow rhombus symbol in Fig. \ref{fig:mass-qf}(b).
 Following the same procedure, mass distributions from QF process for all the incident energies are obtained and shown in Fig. \ref{fig:mass-qf}(b-d). To quantify the peak positions and the width of the peaks for QF modes, all QF mass distributions are fitted using the sum of double Gaussian function. Peak positions and the width of light and heavy fragments are shown in Fig. \ref{fig:mass-qf}(e) and (f) respectively.
It can be observed that for all the beam energies, the mean centroids of the two peaks are located at A=$97 \pm 1$ and A=$160 \pm 1$ as shown in Fig. \ref{fig:mass-qf}(e). It is also interesting to note that the peak heights and hence the contributions from QF process gradually decreases with the increasing beam energies. Both the observations mentioned above hint at possible shell effect in slow QF process.
As already discussed earlier, in a nuclear collision event, when a di-nucleus is formed after penetration of the Coulomb barrier, mass flow can take place from the projectile to the target and vice versa. The former lead to the formation of a compound nucleus followed by the splitting of the nucleus into two fragments. Now if the initial mass-asymmetry is less than the businaro Gallone parameter, the mass flow from the target to the projectile is quite significant. In such cases, it may happen that during the mass equilibration process of the di-nuclear system, when one of the fragment achieves the shell closed configuration, the system gets split into two without driving the system further towards the full mass-equlibriation (compound nuclear formation) process. As soon as the incident energy is increased, either the probability of mass flow from the target to the projectile might be increased or the nucleons participating in the mass-equlibriation process is having more excitation energy which results in reduction of the shell effect in the fragment and hence reduction in the QF probability.


Now the similar analysis has been performed using the mass distribution data for other reaction systems involving heavy projectiles and heavy targets available in the literature. To do so, the measured data available in literature have been compared with the GEF calculations performed by us and from the difference the mass distributions for QF processes have been obtained for different reaction systems at different excitation energies. The experimental FF distributions and the GEF calculations at excitation energy closed to 50 MeV have been shown for the reaction systems $^{9}$Be+$^{249}$Cf\cite{tbanerjee20}, $^{11}$B+$^{238}$U \cite{ss22}, $^{12}$C+$^{238}$U\cite{yadav12}, $^{12}$C+$^{248}$Cm \cite{tbanerjee20},$^{16}$O+ $^{238}$U\cite{tg22}, $^{16}$O+ $^{244}$Pu \cite{tbanerjee20}, $^{18}$O+ $^{232}$Th\cite{yadav12}, $^{24}$Mg+$^{238}$U \cite{hinde18},$^{37}$Cl+$^{206}$Pb \cite{gm20} reaction systems in Fig. \ref{fig:qf-lit}(a-k). The data shown in \ref{fig:qf-lit}(i) for $^{18}$O+ $^{208}$Pb\cite{itkis15} system corresponds to the compound nuclear excitation energy of 87 MeV. The corresponding QF modes for each system (Fig. \ref{fig:qf-lit}(a$'$-k$'$)) derived using the same method have been shown just below the respective mass distribution plots. It can be observed that the QF mass distributions are clearly double humped for all the systems.


To quantify the peak position of the slow QF mode, all QF mass distributions are fitted using the sum of double Gaussian functions as done in Fig. \ref{fig:mass-qf}(b-d) and peak position of the light and heavy fragment are shown as a function of the mass of the di-nuclear system in Fig. \ref{fig:qf-lit}(l). It is very interesting to observe that the peak position corresponding to the light fragment is constant, whereas the mass of the heavy fragment increases with the mass of the composite system. This observation is very analogous to the one for asymmetric fission in actinides where the heavy fragment does not change with the mass of the fissioning nuclei, but the light fragment does \cite{flynn72}. The constancy of the heavy peak in the asymmetric fission of actinides confirmed the role of deformed shell closed nuclei $\sim Z_H=52-56$ in asymmetric fission of actinides \cite{kh00,cb08}. Similarly, the constancy of the lighter peak in the pre-actinide mass region suggested the role of shell closed nuclei $Z_L=34$ in the asymmetric fission of pre-actinides\cite{mahata22}. Using the same analogy, the present observation (constancy of the lighter mass peak) can be treated as clear evidence of shell effect in slow-quasi-fission process. Here the lighter fragments may be the nuclei around $^{96}$Zr, a new doubly magic nucleus \cite{molner90,bobo07}.



Further from the ratio of the QF yield and total fission yield at one particular energy, the QF probability can be obtained. Now the QF probabilities obtained as a function of compound nuclear excitation energy have been shown in Fig. \ref{fig:qf-lit}(m)  by green solid circle for the present system.

It can be observed that the the QF probability goes down with the increasing excitation energy, as qualitatively discussed earlier. It is also interesting to note that the QF probability is considerable even beyond 80 MeV excitation energy. It is worth to mention here that the excitation energy in the di-nuclear system is actually less than the compound nuclear excitation energy as the kinetic energy does not get fully transformed into excitation energy due to incomplete mass-equilibriation, thus resulting in pronounced shell effect in the slow quasi-fission process. It is really hard to calculate actual excitation energy in the di-nuclear system undergoing slow quasi-fission process.

The decrease in pre-equilibrium fission probability with increasing projectile energy has already been established from the several fission fragment angular distribution measurements. An attempt on the determination of slow quasi-fission probability has been made from the measured angular anisotropy\cite{hinde18}. However for the first time mass distributions have been used to obtain the slow QF probabilities as a function of compound nuclear excitation energy and fitted with the function $P_{QF}(E^*)=C*exp(-mE^*)$, where $C$ and $m$ are two fitting constants. From the fitting for the present system, we obtain the parameter values as C=56.2 and m=43.4 MeV$^{-1}$


Now the similar analysis has been performed using the mass distribution data for other reaction systems shown in Fig. \ref{fig:qf-lit}(a-k). The QF probabilities  obtained for the available systems are shown in Fig. \ref{fig:qf-lit}(m) by different symbols with different colors. From the above comparison, it can be observed that the trend in QF probabilities is similar to what we observed for the present system. Now the probabilities are fitted using the same expression $P_{QF}(E^*)=C*exp(-mE^*)$ and shown by the solid line with color same as symbol in Fig. \ref{fig:qf-lit}(m). Here the value for m has been kept same what we obtained for the present system measured over wide excitation energies. It is interesting to note that with the same m value but different C values for different systems, the probabilities could be fitted reasonably well.

Moreover from the Fig. \ref{fig:qf-lit}(n) it can be observed that the C values are increasing with the charge product of target and projectile ($Z_pZ_t$). Now  the C values have been fitted using a linearly increasing function $(C= a+b z_pz_t )$ to have a global picture of the slow-quasi fission probabilities. From the present fitting we obtained $a= 3 \pm 13   $ and $ b= .05 \pm .02 $. Using the empirical formula mentioned above one can determine slow QF probability for any reaction system with certain $Z_pZ_t$ values. However, it is important to mention that the formula does not distinguish reaction systems having same $Z_pZ_t$ value but having different masses of the target and projectile.
However the above empirical formula will be useful to estimate the probability for slow QF process, one of the responsible process for inhibiting the SHE formation. 


In summary, a discrepancy in the measured fragment mass distributions  and calculations incorporating compound nuclear fission process for many reaction systems suggests that the measured data contains contributions not only from the CN-fission process but also from the slow QF-process. So subtraction between the measured data (CN-fission + QF) and the model calculation (CN-fission) results in the distribution one would get only from QF-process. It is interesting to observe that the peak positions of the double humped QF mass-distribution does not change, whereas peak height decreases with the increasing beam energy for all the systems considered. It is more interesting to observe that peak position for the lighter fragment is constant whereas the peak position for the heavier fragment increases with the mass of the di-nuclear system. All of the above observations can be treated as clear evidences of quantum shell effect in slow quasi-fission process. Therefore the present study reveal for the first time the role of a new shell closed nucleus governing the mechanism of slow quasi-fission reaction. Further a formula for slow QF process has been postulated by fitting the QF probabilities for different systems. The present result is expected to help improve the nuclear models in reaching reliable predictive capacities of SHN formation cross-sections. The present letter will also certainly trigger more theoretical and experimental investigations in the field of nuclear reactions, structure and super-heavy element synthesis.

\begin{thebibliography}{62}
\expandafter\ifx\csname natexlab\endcsname\relax\def\natexlab#1{#1}\fi
\expandafter\ifx\csname bibnamefont\endcsname\relax
  \def\bibnamefont#1{#1}\fi
\expandafter\ifx\csname bibfnamefont\endcsname\relax
  \def\bibfnamefont#1{#1}\fi
\expandafter\ifx\csname citenamefont\endcsname\relax
  \def\citenamefont#1{#1}\fi
\expandafter\ifx\csname url\endcsname\relax
  \def\url#1{\texttt{#1}}\fi
\expandafter\ifx\csname urlprefix\endcsname\relax\def\urlprefix{URL }\fi
\providecommand{\bibinfo}[2]{#2}
\providecommand{\eprint}[2][]{\url{#2}}

\bibitem[{\citenamefont{Oganessian et~al.}(2004)\citenamefont{Oganessian,
  Utyonkov, Lobanov, Abdullin, Polyakov, Shirokovsky, Tsyganov, Gulbekian,
  Bogomolov, Gikal et~al.}}]{og04}
\bibinfo{author}{\bibfnamefont{Y.~T.} \bibnamefont{Oganessian}},
  \bibinfo{author}{\bibfnamefont{V.~K.} \bibnamefont{Utyonkov}},
  \bibinfo{author}{\bibfnamefont{Y.~V.} \bibnamefont{Lobanov}},
  \bibinfo{author}{\bibfnamefont{F.~S.} \bibnamefont{Abdullin}},
  \bibinfo{author}{\bibfnamefont{A.~N.} \bibnamefont{Polyakov}},
  \bibinfo{author}{\bibfnamefont{I.~V.} \bibnamefont{Shirokovsky}},
  \bibinfo{author}{\bibfnamefont{Y.~S.} \bibnamefont{Tsyganov}},
  \bibinfo{author}{\bibfnamefont{G.~G.} \bibnamefont{Gulbekian}},
  \bibinfo{author}{\bibfnamefont{S.~L.} \bibnamefont{Bogomolov}},
  \bibinfo{author}{\bibfnamefont{B.~N.} \bibnamefont{Gikal}},
  \bibnamefont{et~al.}, \bibinfo{journal}{Phys. Rev. C}
  \textbf{\bibinfo{volume}{70}}, \bibinfo{pages}{064609}
  (\bibinfo{year}{2004}).

\bibitem[{\citenamefont{Oganessian et~al.}(2006)\citenamefont{Oganessian,
  Utyonkov, Lobanov, Abdullin, Polyakov, Sagaidak, Shirokovsky, Tsyganov,
  Voinov, Gulbekian et~al.}}]{og06}
\bibinfo{author}{\bibfnamefont{Y.~T.} \bibnamefont{Oganessian}},
  \bibinfo{author}{\bibfnamefont{V.~K.} \bibnamefont{Utyonkov}},
  \bibinfo{author}{\bibfnamefont{Y.~V.} \bibnamefont{Lobanov}},
  \bibinfo{author}{\bibfnamefont{F.~S.} \bibnamefont{Abdullin}},
  \bibinfo{author}{\bibfnamefont{A.~N.} \bibnamefont{Polyakov}},
  \bibinfo{author}{\bibfnamefont{R.~N.} \bibnamefont{Sagaidak}},
  \bibinfo{author}{\bibfnamefont{I.~V.} \bibnamefont{Shirokovsky}},
  \bibinfo{author}{\bibfnamefont{Y.~S.} \bibnamefont{Tsyganov}},
  \bibinfo{author}{\bibfnamefont{A.~A.} \bibnamefont{Voinov}},
  \bibinfo{author}{\bibfnamefont{G.~G.} \bibnamefont{Gulbekian}},
  \bibnamefont{et~al.}, \bibinfo{journal}{Phys. Rev. C}
  \textbf{\bibinfo{volume}{74}}, \bibinfo{pages}{044602}
  (\bibinfo{year}{2006}).

\bibitem[{\citenamefont{Ts.Oganessian}(2007)}]{og07}
\bibinfo{author}{\bibfnamefont{Y.}~\bibnamefont{Ts.Oganessian}},
  \bibinfo{journal}{J. Phys. G, Nucl. Part. Phys.}
  \textbf{\bibinfo{volume}{34}}, \bibinfo{pages}{R165} (\bibinfo{year}{2007}).

\bibitem[{\citenamefont{Oganessian et~al.}(2010)\citenamefont{Oganessian,
  Abdullin, Bailey, Benker, Bennett, Dmitriev, Ezold, Hamilton, Henderson,
  Itkis et~al.}}]{og10}
\bibinfo{author}{\bibfnamefont{Y.~T.} \bibnamefont{Oganessian}},
  \bibinfo{author}{\bibfnamefont{F.~S.} \bibnamefont{Abdullin}},
  \bibinfo{author}{\bibfnamefont{P.~D.} \bibnamefont{Bailey}},
  \bibinfo{author}{\bibfnamefont{D.~E.} \bibnamefont{Benker}},
  \bibinfo{author}{\bibfnamefont{M.~E.} \bibnamefont{Bennett}},
  \bibinfo{author}{\bibfnamefont{S.~N.} \bibnamefont{Dmitriev}},
  \bibinfo{author}{\bibfnamefont{J.~G.} \bibnamefont{Ezold}},
  \bibinfo{author}{\bibfnamefont{J.~H.} \bibnamefont{Hamilton}},
  \bibinfo{author}{\bibfnamefont{R.~A.} \bibnamefont{Henderson}},
  \bibinfo{author}{\bibfnamefont{M.~G.} \bibnamefont{Itkis}},
  \bibnamefont{et~al.}, \bibinfo{journal}{Phys. Rev. Lett.}
  \textbf{\bibinfo{volume}{104}}, \bibinfo{pages}{142502}
  (\bibinfo{year}{2010}).

\bibitem[{\citenamefont{Oganessian and Utyonkov}(2015)}]{og15}
\bibinfo{author}{\bibfnamefont{Y.}~\bibnamefont{Oganessian}} \bibnamefont{and}
  \bibinfo{author}{\bibfnamefont{V.}~\bibnamefont{Utyonkov}},
  \bibinfo{journal}{Rep. Prog. Phys.} \textbf{\bibinfo{volume}{78}},
  \bibinfo{pages}{036301} (\bibinfo{year}{2015}).

\bibitem[{\citenamefont{N.Dmitriev et~al.}(2005)\citenamefont{N.Dmitriev,
  Ts.Oganessyan, K.Utyonkov, V.Shishkin, V.Yeremin, V.Lobanov, S.Tsyganov,
  I.Chepygin, A.Sokol, K.Vostokin et~al.}}]{dmitriev05}
\bibinfo{author}{\bibfnamefont{S.}~\bibnamefont{N.Dmitriev}},
  \bibinfo{author}{\bibfnamefont{Y.}~\bibnamefont{Ts.Oganessyan}},
  \bibinfo{author}{\bibfnamefont{V.}~\bibnamefont{K.Utyonkov}},
  \bibinfo{author}{\bibfnamefont{S.}~\bibnamefont{V.Shishkin}},
  \bibinfo{author}{\bibfnamefont{A.}~\bibnamefont{V.Yeremin}},
  \bibinfo{author}{\bibfnamefont{Y.}~\bibnamefont{V.Lobanov}},
  \bibinfo{author}{\bibfnamefont{Y.}~\bibnamefont{S.Tsyganov}},
  \bibinfo{author}{\bibfnamefont{V.}~\bibnamefont{I.Chepygin}},
  \bibinfo{author}{\bibfnamefont{E.}~\bibnamefont{A.Sokol}},
  \bibinfo{author}{\bibfnamefont{G.}~\bibnamefont{K.Vostokin}},
  \bibnamefont{et~al.}, \bibinfo{journal}{Mendeleev Commun.}
  \textbf{\bibinfo{volume}{15}}, \bibinfo{pages}{1} (\bibinfo{year}{2005}).

\bibitem[{\citenamefont{Peter et~al.}(1975)\citenamefont{Peter, Ngo, and
  Tamain}}]{jp75}
\bibinfo{author}{\bibfnamefont{J.}~\bibnamefont{Peter}},
  \bibinfo{author}{\bibfnamefont{C.}~\bibnamefont{Ngo}}, \bibnamefont{and}
  \bibinfo{author}{\bibfnamefont{B.}~\bibnamefont{Tamain}},
  \bibinfo{journal}{Nucl. Phys. A} \textbf{\bibinfo{volume}{250}},
  \bibinfo{pages}{351} (\bibinfo{year}{1975}).

\bibitem[{\citenamefont{Bock et~al.}(1982)\citenamefont{Bock, Chu, Dakowski,
  Gobbi, Grosse, Olmi, Sann, Schwalm, Lynen, Muller et~al.}}]{rb82}
\bibinfo{author}{\bibfnamefont{R.}~\bibnamefont{Bock}},
  \bibinfo{author}{\bibfnamefont{Y.}~\bibnamefont{Chu}},
  \bibinfo{author}{\bibfnamefont{M.}~\bibnamefont{Dakowski}},
  \bibinfo{author}{\bibfnamefont{A.}~\bibnamefont{Gobbi}},
  \bibinfo{author}{\bibfnamefont{E.}~\bibnamefont{Grosse}},
  \bibinfo{author}{\bibfnamefont{A.}~\bibnamefont{Olmi}},
  \bibinfo{author}{\bibfnamefont{H.}~\bibnamefont{Sann}},
  \bibinfo{author}{\bibfnamefont{D.}~\bibnamefont{Schwalm}},
  \bibinfo{author}{\bibfnamefont{U.}~\bibnamefont{Lynen}},
  \bibinfo{author}{\bibfnamefont{W.}~\bibnamefont{Muller}},
  \bibnamefont{et~al.}, \bibinfo{journal}{Nucl. Phys. A}
  \textbf{\bibinfo{volume}{388}}, \bibinfo{pages}{334} (\bibinfo{year}{1982}).

\bibitem[{\citenamefont{Toke et~al.}(1985)\citenamefont{Toke, Bock, Dai, Gobbi,
  Gralla, Hildenbrand, Kuzminski, Muller, Olmi, Stelzer et~al.}}]{jt85}
\bibinfo{author}{\bibfnamefont{J.}~\bibnamefont{Toke}},
  \bibinfo{author}{\bibfnamefont{R.}~\bibnamefont{Bock}},
  \bibinfo{author}{\bibfnamefont{G.}~\bibnamefont{Dai}},
  \bibinfo{author}{\bibfnamefont{A.}~\bibnamefont{Gobbi}},
  \bibinfo{author}{\bibfnamefont{S.}~\bibnamefont{Gralla}},
  \bibinfo{author}{\bibfnamefont{K.}~\bibnamefont{Hildenbrand}},
  \bibinfo{author}{\bibfnamefont{J.}~\bibnamefont{Kuzminski}},
  \bibinfo{author}{\bibfnamefont{W.}~\bibnamefont{Muller}},
  \bibinfo{author}{\bibfnamefont{A.}~\bibnamefont{Olmi}},
  \bibinfo{author}{\bibfnamefont{H.}~\bibnamefont{Stelzer}},
  \bibnamefont{et~al.}, \bibinfo{journal}{Nucl. Phys. A}
  \textbf{\bibinfo{volume}{440}}, \bibinfo{pages}{327} (\bibinfo{year}{1985}).

\bibitem[{\citenamefont{Ramamurthy and Kapoor}(1985)}]{sk85}
\bibinfo{author}{\bibfnamefont{V.~S.} \bibnamefont{Ramamurthy}}
  \bibnamefont{and} \bibinfo{author}{\bibfnamefont{S.~S.}
  \bibnamefont{Kapoor}}, \bibinfo{journal}{Phys. Rev. Lett}
  \textbf{\bibinfo{volume}{54}}, \bibinfo{pages}{178} (\bibinfo{year}{1985}).

\bibitem[{\citenamefont{Hinde et~al.}(1995)\citenamefont{Hinde, Dasgupta,
  Leigh, Lestone, Mein, Morton, Newton, and Timmers}}]{hinde95}
\bibinfo{author}{\bibfnamefont{D.~J.} \bibnamefont{Hinde}},
  \bibinfo{author}{\bibfnamefont{M.}~\bibnamefont{Dasgupta}},
  \bibinfo{author}{\bibfnamefont{J.~R.} \bibnamefont{Leigh}},
  \bibinfo{author}{\bibfnamefont{J.~P.} \bibnamefont{Lestone}},
  \bibinfo{author}{\bibfnamefont{J.~C.} \bibnamefont{Mein}},
  \bibinfo{author}{\bibfnamefont{C.~R.} \bibnamefont{Morton}},
  \bibinfo{author}{\bibfnamefont{J.~O.} \bibnamefont{Newton}},
  \bibnamefont{and} \bibinfo{author}{\bibfnamefont{H.}~\bibnamefont{Timmers}},
  \bibinfo{journal}{Phys. Rev. Lett.} \textbf{\bibinfo{volume}{74}},
  \bibinfo{pages}{1295} (\bibinfo{year}{1995}).

\bibitem[{\citenamefont{Ramamurthy et~al.}(1990)\citenamefont{Ramamurthy,
  Kapoor, Choudhury, Saxena, Nadkarni, Mohanty, Nayak, Sastry, Kailas,
  Chatterjee et~al.}}]{rama90}
\bibinfo{author}{\bibfnamefont{V.~S.} \bibnamefont{Ramamurthy}},
  \bibinfo{author}{\bibfnamefont{S.~S.} \bibnamefont{Kapoor}},
  \bibinfo{author}{\bibfnamefont{R.~K.} \bibnamefont{Choudhury}},
  \bibinfo{author}{\bibfnamefont{A.}~\bibnamefont{Saxena}},
  \bibinfo{author}{\bibfnamefont{D.~M.} \bibnamefont{Nadkarni}},
  \bibinfo{author}{\bibfnamefont{A.~K.} \bibnamefont{Mohanty}},
  \bibinfo{author}{\bibfnamefont{B.~K.} \bibnamefont{Nayak}},
  \bibinfo{author}{\bibfnamefont{S.~V.} \bibnamefont{Sastry}},
  \bibinfo{author}{\bibfnamefont{S.}~\bibnamefont{Kailas}},
  \bibinfo{author}{\bibfnamefont{A.}~\bibnamefont{Chatterjee}},
  \bibnamefont{et~al.}, \bibinfo{journal}{Phys. Rev. Lett.}
  \textbf{\bibinfo{volume}{65}}, \bibinfo{pages}{25} (\bibinfo{year}{1990}).

\bibitem[{\citenamefont{D.J.Hinde et~al.}(2008)\citenamefont{D.J.Hinde, Rietz,
  M.Dasgupta, R.G.Thomas, and L.R.Gasques}}]{hinde08}
\bibinfo{author}{\bibnamefont{D.J.Hinde}},
  \bibinfo{author}{\bibfnamefont{R.}~\bibnamefont{Rietz}},
  \bibinfo{author}{\bibnamefont{M.Dasgupta}},
  \bibinfo{author}{\bibnamefont{R.G.Thomas}}, \bibnamefont{and}
  \bibinfo{author}{\bibnamefont{L.R.Gasques}}, \bibinfo{journal}{Phys. Rev.
  Lett.} \textbf{\bibinfo{volume}{101}}, \bibinfo{pages}{092701}
  (\bibinfo{year}{2008}).

\bibitem[{\citenamefont{D.J.Hinde et~al.}(2018)\citenamefont{D.J.Hinde,
  D.Y.Jeung, E.Prasad, A.Wakhle, M.Dasgupta, M.Evers, D.H.Luong, Rietz,
  C.Simenel, E.C.Simpson et~al.}}]{hinde18}
\bibinfo{author}{\bibnamefont{D.J.Hinde}},
  \bibinfo{author}{\bibnamefont{D.Y.Jeung}},
  \bibinfo{author}{\bibnamefont{E.Prasad}},
  \bibinfo{author}{\bibnamefont{A.Wakhle}},
  \bibinfo{author}{\bibnamefont{M.Dasgupta}},
  \bibinfo{author}{\bibnamefont{M.Evers}},
  \bibinfo{author}{\bibnamefont{D.H.Luong}},
  \bibinfo{author}{\bibfnamefont{R.}~\bibnamefont{Rietz}},
  \bibinfo{author}{\bibnamefont{C.Simenel}},
  \bibinfo{author}{\bibnamefont{E.C.Simpson}}, \bibnamefont{et~al.},
  \bibinfo{journal}{Phys. Rev. C} \textbf{\bibinfo{volume}{97}},
  \bibinfo{pages}{024616} (\bibinfo{year}{2018}).

\bibitem[{\citenamefont{Rafiei et~al.}(2008)\citenamefont{Rafiei, Thomas,
  Hinde, Dasgupta, Morton, Gasques, Brown, and Rodriguez}}]{rafiei08}
\bibinfo{author}{\bibfnamefont{R.}~\bibnamefont{Rafiei}},
  \bibinfo{author}{\bibfnamefont{R.~G.} \bibnamefont{Thomas}},
  \bibinfo{author}{\bibfnamefont{D.~J.} \bibnamefont{Hinde}},
  \bibinfo{author}{\bibfnamefont{M.}~\bibnamefont{Dasgupta}},
  \bibinfo{author}{\bibfnamefont{C.~R.} \bibnamefont{Morton}},
  \bibinfo{author}{\bibfnamefont{L.~R.} \bibnamefont{Gasques}},
  \bibinfo{author}{\bibfnamefont{M.~L.} \bibnamefont{Brown}}, \bibnamefont{and}
  \bibinfo{author}{\bibfnamefont{M.~D.} \bibnamefont{Rodriguez}},
  \bibinfo{journal}{Phys. Rev. C} \textbf{\bibinfo{volume}{77}},
  \bibinfo{pages}{024606} (\bibinfo{year}{2008}).

\bibitem[{\citenamefont{E.Prasad et~al.}(2016)\citenamefont{E.Prasad, A.Wakhle,
  D.J.Hinde, E.Williams, M.Dasgupta, M.Evers, D.H.Luong, G.Mohanto, C.Simenel,
  and K.Vo-Phuoc}}]{prasad16}
\bibinfo{author}{\bibnamefont{E.Prasad}},
  \bibinfo{author}{\bibnamefont{A.Wakhle}},
  \bibinfo{author}{\bibnamefont{D.J.Hinde}},
  \bibinfo{author}{\bibnamefont{E.Williams}},
  \bibinfo{author}{\bibnamefont{M.Dasgupta}},
  \bibinfo{author}{\bibnamefont{M.Evers}},
  \bibinfo{author}{\bibnamefont{D.H.Luong}},
  \bibinfo{author}{\bibnamefont{G.Mohanto}},
  \bibinfo{author}{\bibnamefont{C.Simenel}}, \bibnamefont{and}
  \bibinfo{author}{\bibnamefont{K.Vo-Phuoc}}, \bibinfo{journal}{Phys. Rev. C}
  \textbf{\bibinfo{volume}{93}}, \bibinfo{pages}{024607}
  (\bibinfo{year}{2016}).

\bibitem[{\citenamefont{Kozulin et~al.}(2016)\citenamefont{Kozulin, Knyazheva,
  Novikov, Itkis, Itkis, Dmitriev, Oganessian, Bogachev, Kozulina, Harca
  et~al.}}]{kozulin16}
\bibinfo{author}{\bibfnamefont{E.~M.} \bibnamefont{Kozulin}},
  \bibinfo{author}{\bibfnamefont{G.~N.} \bibnamefont{Knyazheva}},
  \bibinfo{author}{\bibfnamefont{K.~V.} \bibnamefont{Novikov}},
  \bibinfo{author}{\bibfnamefont{I.~M.} \bibnamefont{Itkis}},
  \bibinfo{author}{\bibfnamefont{M.~G.} \bibnamefont{Itkis}},
  \bibinfo{author}{\bibfnamefont{S.~N.} \bibnamefont{Dmitriev}},
  \bibinfo{author}{\bibfnamefont{Y.~T.} \bibnamefont{Oganessian}},
  \bibinfo{author}{\bibfnamefont{A.~A.} \bibnamefont{Bogachev}},
  \bibinfo{author}{\bibfnamefont{N.~I.} \bibnamefont{Kozulina}},
  \bibinfo{author}{\bibfnamefont{I.}~\bibnamefont{Harca}},
  \bibnamefont{et~al.}, \bibinfo{journal}{Phys. Rev. C}
  \textbf{\bibinfo{volume}{94}}, \bibinfo{pages}{054613}
  (\bibinfo{year}{2016}).

\bibitem[{\citenamefont{Itkis et~al.}(2015)\citenamefont{Itkis, Vardaci, Itkis,
  Knyazheva, and Kozulin}}]{itkis15}
\bibinfo{author}{\bibfnamefont{M.~G.} \bibnamefont{Itkis}},
  \bibinfo{author}{\bibfnamefont{E.}~\bibnamefont{Vardaci}},
  \bibinfo{author}{\bibfnamefont{I.~M.} \bibnamefont{Itkis}},
  \bibinfo{author}{\bibfnamefont{G.~N.} \bibnamefont{Knyazheva}},
  \bibnamefont{and} \bibinfo{author}{\bibfnamefont{E.~M.}
  \bibnamefont{Kozulin}}, \bibinfo{journal}{Nucl. Phys. A}
  \textbf{\bibinfo{volume}{944}}, \bibinfo{pages}{204} (\bibinfo{year}{2015}).

\bibitem[{\citenamefont{A.Sen et~al.}(2022)\citenamefont{A.Sen, T.K.Ghosh,
  E.M.Kozulin, I.M.Itkis, G.N.Knyazheva, K.V.Novikov, S.Bhattacharya,
  K.Banerjee, and C.Bhattacharya}}]{asen22}
\bibinfo{author}{\bibnamefont{A.Sen}},
  \bibinfo{author}{\bibnamefont{T.K.Ghosh}},
  \bibinfo{author}{\bibnamefont{E.M.Kozulin}},
  \bibinfo{author}{\bibnamefont{I.M.Itkis}},
  \bibinfo{author}{\bibnamefont{G.N.Knyazheva}},
  \bibinfo{author}{\bibnamefont{K.V.Novikov}},
  \bibinfo{author}{\bibnamefont{S.Bhattacharya}},
  \bibinfo{author}{\bibnamefont{K.Banerjee}}, \bibnamefont{and}
  \bibinfo{author}{\bibnamefont{C.Bhattacharya}}, \bibinfo{journal}{Phys. Rev.
  C} \textbf{\bibinfo{volume}{105}}, \bibinfo{pages}{014627}
  (\bibinfo{year}{2022}).

\bibitem[{\citenamefont{Thomas et~al.}(2008)\citenamefont{Thomas, Hinde,
  Duniec, Zenke, Dasgupta, Brown, Evers, Gasques, Rodriguez, and
  Diaz-Torres}}]{thomas08}
\bibinfo{author}{\bibfnamefont{R.~G.} \bibnamefont{Thomas}},
  \bibinfo{author}{\bibfnamefont{D.~J.} \bibnamefont{Hinde}},
  \bibinfo{author}{\bibfnamefont{D.}~\bibnamefont{Duniec}},
  \bibinfo{author}{\bibfnamefont{F.}~\bibnamefont{Zenke}},
  \bibinfo{author}{\bibfnamefont{M.}~\bibnamefont{Dasgupta}},
  \bibinfo{author}{\bibfnamefont{M.~L.} \bibnamefont{Brown}},
  \bibinfo{author}{\bibfnamefont{M.}~\bibnamefont{Evers}},
  \bibinfo{author}{\bibfnamefont{L.~R.} \bibnamefont{Gasques}},
  \bibinfo{author}{\bibfnamefont{M.~D.} \bibnamefont{Rodriguez}},
  \bibnamefont{and}
  \bibinfo{author}{\bibfnamefont{A.}~\bibnamefont{Diaz-Torres}},
  \bibinfo{journal}{Phys. Rev. C} \textbf{\bibinfo{volume}{77}},
  \bibinfo{pages}{034610} (\bibinfo{year}{2008}).

\bibitem[{\citenamefont{C.Yadav et~al.}(2012)\citenamefont{C.Yadav, R.G.Thomas,
  R.K.Choudhury, P.Sugathan, A.Jhingan, S.Appannababu, K.S.Golda, D.Singh,
  Mukul, J.Gehlot et~al.}}]{yadav12}
\bibinfo{author}{\bibnamefont{C.Yadav}},
  \bibinfo{author}{\bibnamefont{R.G.Thomas}},
  \bibinfo{author}{\bibnamefont{R.K.Choudhury}},
  \bibinfo{author}{\bibnamefont{P.Sugathan}},
  \bibinfo{author}{\bibnamefont{A.Jhingan}},
  \bibinfo{author}{\bibnamefont{S.Appannababu}},
  \bibinfo{author}{\bibnamefont{K.S.Golda}},
  \bibinfo{author}{\bibnamefont{D.Singh}},
  \bibinfo{author}{\bibfnamefont{I.}~\bibnamefont{Mukul}},
  \bibinfo{author}{\bibnamefont{J.Gehlot}}, \bibnamefont{et~al.},
  \bibinfo{journal}{Phys. Rev. C} \textbf{\bibinfo{volume}{86}},
  \bibinfo{pages}{034606} (\bibinfo{year}{2012}).

\bibitem[{\citenamefont{T.Banerjee et~al.}(2020)\citenamefont{T.Banerjee,
  D.J.Hinde, and et~al.}}]{tbanerjee20}
\bibinfo{author}{\bibnamefont{T.Banerjee}},
  \bibinfo{author}{\bibnamefont{D.J.Hinde}}, \bibnamefont{and}
  \bibinfo{author}{\bibnamefont{et~al.}}, \bibinfo{journal}{Phys. Rev. C}
  \textbf{\bibinfo{volume}{102}}, \bibinfo{pages}{024603}
  (\bibinfo{year}{2020}).

\bibitem[{\citenamefont{K.Banerjee et~al.}(2011)\citenamefont{K.Banerjee,
  T.K.Ghosh, S.Bhattacharya, C.Bhattacharya, S.Kundu, T.K.Rana, G.Mukherjee,
  J.K.Meena, J.Sadhukhan, S.Pal et~al.}}]{banerjee11}
\bibinfo{author}{\bibnamefont{K.Banerjee}},
  \bibinfo{author}{\bibnamefont{T.K.Ghosh}},
  \bibinfo{author}{\bibnamefont{S.Bhattacharya}},
  \bibinfo{author}{\bibnamefont{C.Bhattacharya}},
  \bibinfo{author}{\bibnamefont{S.Kundu}},
  \bibinfo{author}{\bibnamefont{T.K.Rana}},
  \bibinfo{author}{\bibnamefont{G.Mukherjee}},
  \bibinfo{author}{\bibnamefont{J.K.Meena}},
  \bibinfo{author}{\bibnamefont{J.Sadhukhan}},
  \bibinfo{author}{\bibnamefont{S.Pal}}, \bibnamefont{et~al.},
  \bibinfo{journal}{Phys. Rev. C} \textbf{\bibinfo{volume}{83}},
  \bibinfo{pages}{024605} (\bibinfo{year}{2011}).

\bibitem[{\citenamefont{Ghosh et~al.}(2004{\natexlab{a}})\citenamefont{Ghosh,
  Pal, Sinha, Majumdar, Chattopadhyay, Bhattacharya, Saxena, Sahu, Golda, and
  Datta}}]{ghosh05}
\bibinfo{author}{\bibfnamefont{T.~K.} \bibnamefont{Ghosh}},
  \bibinfo{author}{\bibfnamefont{S.}~\bibnamefont{Pal}},
  \bibinfo{author}{\bibfnamefont{T.}~\bibnamefont{Sinha}},
  \bibinfo{author}{\bibfnamefont{N.}~\bibnamefont{Majumdar}},
  \bibinfo{author}{\bibfnamefont{S.}~\bibnamefont{Chattopadhyay}},
  \bibinfo{author}{\bibfnamefont{P.}~\bibnamefont{Bhattacharya}},
  \bibinfo{author}{\bibfnamefont{A.}~\bibnamefont{Saxena}},
  \bibinfo{author}{\bibfnamefont{P.~K.} \bibnamefont{Sahu}},
  \bibinfo{author}{\bibfnamefont{K.~S.} \bibnamefont{Golda}}, \bibnamefont{and}
  \bibinfo{author}{\bibfnamefont{S.~K.} \bibnamefont{Datta}},
  \bibinfo{journal}{Phys. Rev. C} \textbf{\bibinfo{volume}{69}},
  \bibinfo{pages}{031603(R)} (\bibinfo{year}{2004}{\natexlab{a}}).

\bibitem[{\citenamefont{Ghosh et~al.}(2009)\citenamefont{Ghosh, Banerjee,
  Bhattacharya, Bhattacharya, Kundu, Mali, Meena, Mukherjee, Mukhopadhyay, Rana
  et~al.}}]{ghosh09}
\bibinfo{author}{\bibfnamefont{T.~K.} \bibnamefont{Ghosh}},
  \bibinfo{author}{\bibfnamefont{K.}~\bibnamefont{Banerjee}},
  \bibinfo{author}{\bibfnamefont{C.}~\bibnamefont{Bhattacharya}},
  \bibinfo{author}{\bibfnamefont{S.}~\bibnamefont{Bhattacharya}},
  \bibinfo{author}{\bibfnamefont{S.}~\bibnamefont{Kundu}},
  \bibinfo{author}{\bibfnamefont{P.}~\bibnamefont{Mali}},
  \bibinfo{author}{\bibfnamefont{J.~K.} \bibnamefont{Meena}},
  \bibinfo{author}{\bibfnamefont{G.}~\bibnamefont{Mukherjee}},
  \bibinfo{author}{\bibfnamefont{S.}~\bibnamefont{Mukhopadhyay}},
  \bibinfo{author}{\bibfnamefont{T.~K.} \bibnamefont{Rana}},
  \bibnamefont{et~al.}, \bibinfo{journal}{Phys. Rev. C}
  \textbf{\bibinfo{volume}{79}}, \bibinfo{pages}{054607}
  (\bibinfo{year}{2009}).

\bibitem[{\citenamefont{Ghosh et~al.}(2004{\natexlab{b}})\citenamefont{Ghosh,
  Pal, Sinha, Chattopadhyay, Bhattacharya, Biswas, and Golda}}]{ghosh041}
\bibinfo{author}{\bibfnamefont{T.~K.} \bibnamefont{Ghosh}},
  \bibinfo{author}{\bibfnamefont{S.}~\bibnamefont{Pal}},
  \bibinfo{author}{\bibfnamefont{T.}~\bibnamefont{Sinha}},
  \bibinfo{author}{\bibfnamefont{S.}~\bibnamefont{Chattopadhyay}},
  \bibinfo{author}{\bibfnamefont{P.}~\bibnamefont{Bhattacharya}},
  \bibinfo{author}{\bibfnamefont{D.~C.} \bibnamefont{Biswas}},
  \bibnamefont{and} \bibinfo{author}{\bibfnamefont{K.~S.} \bibnamefont{Golda}},
  \bibinfo{journal}{Phys. Rev. C} \textbf{\bibinfo{volume}{70}},
  \bibinfo{pages}{011604(R)} (\bibinfo{year}{2004}{\natexlab{b}}).

\bibitem[{\citenamefont{M.G.Itkis et~al.}(2007)\citenamefont{M.G.Itkis,
  A.A.Bogachev, I.M.Itkis, J.Kliman, G.N.Knyazheva, N.A.Kondratiev,
  E.M.Kozulin, L.Krupa, Yu.Ts.Oganessian, I.V.Pokrovsky et~al.}}]{mgitkis07}
\bibinfo{author}{\bibnamefont{M.G.Itkis}},
  \bibinfo{author}{\bibnamefont{A.A.Bogachev}},
  \bibinfo{author}{\bibnamefont{I.M.Itkis}},
  \bibinfo{author}{\bibnamefont{J.Kliman}},
  \bibinfo{author}{\bibnamefont{G.N.Knyazheva}},
  \bibinfo{author}{\bibnamefont{N.A.Kondratiev}},
  \bibinfo{author}{\bibnamefont{E.M.Kozulin}},
  \bibinfo{author}{\bibnamefont{L.Krupa}},
  \bibinfo{author}{\bibnamefont{Yu.Ts.Oganessian}},
  \bibinfo{author}{\bibnamefont{I.V.Pokrovsky}}, \bibnamefont{et~al.},
  \bibinfo{journal}{Nucl. Phys. A} \textbf{\bibinfo{volume}{787}},
  \bibinfo{pages}{150c} (\bibinfo{year}{2007}).

\bibitem[{\citenamefont{Morjean et~al.}(2008)\citenamefont{Morjean, Jacquet,
  Charvet, L Hoir, Laget, Parlog, Chbihi, Chevallier, Cohen, Dauvergne
  et~al.}}]{morjean08}
\bibinfo{author}{\bibfnamefont{M.}~\bibnamefont{Morjean}},
  \bibinfo{author}{\bibfnamefont{D.}~\bibnamefont{Jacquet}},
  \bibinfo{author}{\bibfnamefont{J.~L.} \bibnamefont{Charvet}},
  \bibinfo{author}{\bibfnamefont{A.}~\bibnamefont{L Hoir}},
  \bibinfo{author}{\bibfnamefont{M.}~\bibnamefont{Laget}},
  \bibinfo{author}{\bibfnamefont{M.}~\bibnamefont{Parlog}},
  \bibinfo{author}{\bibfnamefont{A.}~\bibnamefont{Chbihi}},
  \bibinfo{author}{\bibfnamefont{M.}~\bibnamefont{Chevallier}},
  \bibinfo{author}{\bibfnamefont{C.}~\bibnamefont{Cohen}},
  \bibinfo{author}{\bibfnamefont{D.}~\bibnamefont{Dauvergne}},
  \bibnamefont{et~al.}, \bibinfo{journal}{Phys. Rev. Lett}
  \textbf{\bibinfo{volume}{101}}, \bibinfo{pages}{072701}
  (\bibinfo{year}{2008}).

\bibitem[{\citenamefont{Diaz-Torres et~al.}(2001)\citenamefont{Diaz-Torres,
  Adamian, Antonenko, and Scheid}}]{torres01}
\bibinfo{author}{\bibfnamefont{A.}~\bibnamefont{Diaz-Torres}},
  \bibinfo{author}{\bibfnamefont{G.~G.} \bibnamefont{Adamian}},
  \bibinfo{author}{\bibfnamefont{N.~V.} \bibnamefont{Antonenko}},
  \bibnamefont{and} \bibinfo{author}{\bibfnamefont{W.}~\bibnamefont{Scheid}},
  \bibinfo{journal}{Phys. Rev. C} \textbf{\bibinfo{volume}{64}},
  \bibinfo{pages}{024604} (\bibinfo{year}{2001}).

\bibitem[{\citenamefont{Aritomo et~al.}(2012)\citenamefont{Aritomo, Hagino,
  Nishio, and Chiba}}]{aritomo12}
\bibinfo{author}{\bibfnamefont{Y.}~\bibnamefont{Aritomo}},
  \bibinfo{author}{\bibfnamefont{K.}~\bibnamefont{Hagino}},
  \bibinfo{author}{\bibfnamefont{K.}~\bibnamefont{Nishio}}, \bibnamefont{and}
  \bibinfo{author}{\bibfnamefont{S.}~\bibnamefont{Chiba}},
  \bibinfo{journal}{Phys. Rev. C} \textbf{\bibinfo{volume}{85}},
  \bibinfo{pages}{044614} (\bibinfo{year}{2012}).

\bibitem[{\citenamefont{Simenel}(2012)}]{sim12}
\bibinfo{author}{\bibfnamefont{C.}~\bibnamefont{Simenel}},
  \bibinfo{journal}{Eur. Phys. J. A} \textbf{\bibinfo{volume}{48}},
  \bibinfo{pages}{152} (\bibinfo{year}{2012}).

\bibitem[{\citenamefont{Simenel et~al.}(2012)\citenamefont{Simenel, Hinde,
  du~Rietz, Dasgupta, Evers, Lin, Luong, and Wakhle}}]{simenel12}
\bibinfo{author}{\bibfnamefont{C.}~\bibnamefont{Simenel}},
  \bibinfo{author}{\bibfnamefont{D.}~\bibnamefont{Hinde}},
  \bibinfo{author}{\bibfnamefont{R.}~\bibnamefont{du~Rietz}},
  \bibinfo{author}{\bibfnamefont{M.}~\bibnamefont{Dasgupta}},
  \bibinfo{author}{\bibfnamefont{M.}~\bibnamefont{Evers}},
  \bibinfo{author}{\bibfnamefont{C.}~\bibnamefont{Lin}},
  \bibinfo{author}{\bibfnamefont{D.}~\bibnamefont{Luong}}, \bibnamefont{and}
  \bibinfo{author}{\bibfnamefont{A.}~\bibnamefont{Wakhle}},
  \bibinfo{journal}{Phys. Lett. B} \textbf{\bibinfo{volume}{710}},
  \bibinfo{pages}{607} (\bibinfo{year}{2012}).

\bibitem[{\citenamefont{Sekizawa and Yabana}(2016)}]{seki16}
\bibinfo{author}{\bibfnamefont{K.}~\bibnamefont{Sekizawa}} \bibnamefont{and}
  \bibinfo{author}{\bibfnamefont{K.}~\bibnamefont{Yabana}},
  \bibinfo{journal}{Phys. Rev. C} \textbf{\bibinfo{volume}{93}},
  \bibinfo{pages}{054616} (\bibinfo{year}{2016}).

\bibitem[{\citenamefont{M.G.Itkis et~al.}(2004)\citenamefont{M.G.Itkis,
  J.Aysto, S.Beghini, A.A.Bogachev, L.Corradi, O.Dorvaux, A.Gadea, G.Giardina,
  F.Hanappe, I.M.Itkis et~al.}}]{mitkis04}
\bibinfo{author}{\bibnamefont{M.G.Itkis}},
  \bibinfo{author}{\bibnamefont{J.Aysto}},
  \bibinfo{author}{\bibnamefont{S.Beghini}},
  \bibinfo{author}{\bibnamefont{A.A.Bogachev}},
  \bibinfo{author}{\bibnamefont{L.Corradi}},
  \bibinfo{author}{\bibnamefont{O.Dorvaux}},
  \bibinfo{author}{\bibnamefont{A.Gadea}},
  \bibinfo{author}{\bibnamefont{G.Giardina}},
  \bibinfo{author}{\bibnamefont{F.Hanappe}},
  \bibinfo{author}{\bibnamefont{I.M.Itkis}}, \bibnamefont{et~al.},
  \bibinfo{journal}{Nucl. Phys. A} \textbf{\bibinfo{volume}{734}},
  \bibinfo{pages}{136} (\bibinfo{year}{2004}).

\bibitem[{\citenamefont{Wakhle et~al.}(2014)\citenamefont{Wakhle, Simenel,
  Hinde, Dasgupta, Evers, Luong, duRietz, and Williams}}]{wakhle14}
\bibinfo{author}{\bibfnamefont{A.}~\bibnamefont{Wakhle}},
  \bibinfo{author}{\bibfnamefont{C.}~\bibnamefont{Simenel}},
  \bibinfo{author}{\bibfnamefont{D.}~\bibnamefont{Hinde}},
  \bibinfo{author}{\bibfnamefont{M.}~\bibnamefont{Dasgupta}},
  \bibinfo{author}{\bibfnamefont{M.}~\bibnamefont{Evers}},
  \bibinfo{author}{\bibfnamefont{D.}~\bibnamefont{Luong}},
  \bibinfo{author}{\bibfnamefont{R.}~\bibnamefont{duRietz}}, \bibnamefont{and}
  \bibinfo{author}{\bibfnamefont{E.}~\bibnamefont{Williams}},
  \bibinfo{journal}{Phys. Rev. Lett.} \textbf{\bibinfo{volume}{113}},
  \bibinfo{pages}{182502} (\bibinfo{year}{2014}).

\bibitem[{\citenamefont{Oberacker et~al.}(2014)\citenamefont{Oberacker, Umar,
  and Simenel}}]{oberacker14}
\bibinfo{author}{\bibfnamefont{V.}~\bibnamefont{Oberacker}},
  \bibinfo{author}{\bibfnamefont{A.}~\bibnamefont{Umar}}, \bibnamefont{and}
  \bibinfo{author}{\bibfnamefont{C.}~\bibnamefont{Simenel}},
  \bibinfo{journal}{Phys. Rev. C} \textbf{\bibinfo{volume}{90}},
  \bibinfo{pages}{054605} (\bibinfo{year}{2014}).

\bibitem[{\citenamefont{Simenel et~al.}(2021)\citenamefont{Simenel, McGlynn,
  Umar, and Godbey}}]{simenel21}
\bibinfo{author}{\bibfnamefont{C.}~\bibnamefont{Simenel}},
  \bibinfo{author}{\bibfnamefont{P.}~\bibnamefont{McGlynn}},
  \bibinfo{author}{\bibfnamefont{A.}~\bibnamefont{Umar}}, \bibnamefont{and}
  \bibinfo{author}{\bibfnamefont{K.}~\bibnamefont{Godbey}},
  \bibinfo{journal}{Phys. Lett. B} \textbf{\bibinfo{volume}{822}},
  \bibinfo{pages}{136648} (\bibinfo{year}{2021}).

\bibitem[{\citenamefont{Morjean et~al.}(2017)\citenamefont{Morjean, Hinde,
  Simenel, Jeung, Airiau, Cook, Dasgupta, Drouart, Jacquet, Kalkal
  et~al.}}]{morjean17}
\bibinfo{author}{\bibfnamefont{M.}~\bibnamefont{Morjean}},
  \bibinfo{author}{\bibfnamefont{D.}~\bibnamefont{Hinde}},
  \bibinfo{author}{\bibfnamefont{C.}~\bibnamefont{Simenel}},
  \bibinfo{author}{\bibfnamefont{D.}~\bibnamefont{Jeung}},
  \bibinfo{author}{\bibfnamefont{M.}~\bibnamefont{Airiau}},
  \bibinfo{author}{\bibfnamefont{K.}~\bibnamefont{Cook}},
  \bibinfo{author}{\bibfnamefont{M.}~\bibnamefont{Dasgupta}},
  \bibinfo{author}{\bibfnamefont{A.}~\bibnamefont{Drouart}},
  \bibinfo{author}{\bibfnamefont{D.}~\bibnamefont{Jacquet}},
  \bibinfo{author}{\bibfnamefont{S.}~\bibnamefont{Kalkal}},
  \bibnamefont{et~al.}, \bibinfo{journal}{Phys. Rev. Lett.}
  \textbf{\bibinfo{volume}{119}}, \bibinfo{pages}{222502}
  (\bibinfo{year}{2017}).

\bibitem[{\citenamefont{Bockstiegel et~al.}(2008)\citenamefont{Bockstiegel,
  Steinhauser, Schmidt, Clerc, Grewe, Heinz, de~Jong, Junghans, Muller, and
  Voss}}]{cb08}
\bibinfo{author}{\bibfnamefont{C.}~\bibnamefont{Bockstiegel}},
  \bibinfo{author}{\bibfnamefont{S.}~\bibnamefont{Steinhauser}},
  \bibinfo{author}{\bibfnamefont{K.~H.} \bibnamefont{Schmidt}},
  \bibinfo{author}{\bibfnamefont{H.-G.} \bibnamefont{Clerc}},
  \bibinfo{author}{\bibfnamefont{A.}~\bibnamefont{Grewe}},
  \bibinfo{author}{\bibfnamefont{A.}~\bibnamefont{Heinz}},
  \bibinfo{author}{\bibfnamefont{M.}~\bibnamefont{de~Jong}},
  \bibinfo{author}{\bibfnamefont{A.~R.} \bibnamefont{Junghans}},
  \bibinfo{author}{\bibfnamefont{J.}~\bibnamefont{Muller}}, \bibnamefont{and}
  \bibinfo{author}{\bibfnamefont{B.}~\bibnamefont{Voss}},
  \bibinfo{journal}{Nucl. Phys. A} \textbf{\bibinfo{volume}{802}},
  \bibinfo{pages}{12} (\bibinfo{year}{2008}).

\bibitem[{\citenamefont{Scamps and Simenel}(2018)}]{gc18}
\bibinfo{author}{\bibfnamefont{G.}~\bibnamefont{Scamps}} \bibnamefont{and}
  \bibinfo{author}{\bibfnamefont{C.}~\bibnamefont{Simenel}},
  \bibinfo{journal}{Nature (London)} \textbf{\bibinfo{volume}{564}},
  \bibinfo{pages}{382} (\bibinfo{year}{2018}).

\bibitem[{\citenamefont{Santra et~al.}(2014)\citenamefont{Santra, Pal, Rath,
  Nayak, Singh, Chattopadhyay, Behera, Singh, Jhingan, Sugathan
  et~al.}}]{santra14}
\bibinfo{author}{\bibfnamefont{S.}~\bibnamefont{Santra}},
  \bibinfo{author}{\bibfnamefont{A.}~\bibnamefont{Pal}},
  \bibinfo{author}{\bibfnamefont{P.~K.} \bibnamefont{Rath}},
  \bibinfo{author}{\bibfnamefont{B.~K.} \bibnamefont{Nayak}},
  \bibinfo{author}{\bibfnamefont{N.~L.} \bibnamefont{Singh}},
  \bibinfo{author}{\bibfnamefont{D.}~\bibnamefont{Chattopadhyay}},
  \bibinfo{author}{\bibfnamefont{B.~R.} \bibnamefont{Behera}},
  \bibinfo{author}{\bibfnamefont{V.}~\bibnamefont{Singh}},
  \bibinfo{author}{\bibfnamefont{A.}~\bibnamefont{Jhingan}},
  \bibinfo{author}{\bibfnamefont{P.}~\bibnamefont{Sugathan}},
  \bibnamefont{et~al.}, \bibinfo{journal}{Phys. Rev. C}
  \textbf{\bibinfo{volume}{90}}, \bibinfo{pages}{064620}
  (\bibinfo{year}{2014}).

\bibitem[{\citenamefont{Pal et~al.}(2018)\citenamefont{Pal, Santra,
  Chattopadhyay, Kundu, Jhingan, Sugathan, Saneesh, Kumar, N.L.Singh, Yadav
  et~al.}}]{pal18}
\bibinfo{author}{\bibfnamefont{A.}~\bibnamefont{Pal}},
  \bibinfo{author}{\bibfnamefont{S.}~\bibnamefont{Santra}},
  \bibinfo{author}{\bibfnamefont{D.}~\bibnamefont{Chattopadhyay}},
  \bibinfo{author}{\bibfnamefont{A.}~\bibnamefont{Kundu}},
  \bibinfo{author}{\bibfnamefont{A.}~\bibnamefont{Jhingan}},
  \bibinfo{author}{\bibfnamefont{P.}~\bibnamefont{Sugathan}},
  \bibinfo{author}{\bibfnamefont{N.}~\bibnamefont{Saneesh}},
  \bibinfo{author}{\bibfnamefont{M.}~\bibnamefont{Kumar}},
  \bibinfo{author}{\bibnamefont{N.L.Singh}},
  \bibinfo{author}{\bibfnamefont{A.}~\bibnamefont{Yadav}},
  \bibnamefont{et~al.}, \bibinfo{journal}{Phys. Rev. C}
  \textbf{\bibinfo{volume}{98}}, \bibinfo{pages}{031601(R)}
  (\bibinfo{year}{2018}).

\bibitem[{\citenamefont{Pal et~al.}(2021)\citenamefont{Pal, Santra, Rout,
  Gandhi, Baishya, Santhosh, Tripathi, and Nag}}]{apal21}
\bibinfo{author}{\bibfnamefont{A.}~\bibnamefont{Pal}},
  \bibinfo{author}{\bibfnamefont{S.}~\bibnamefont{Santra}},
  \bibinfo{author}{\bibfnamefont{P.}~\bibnamefont{Rout}},
  \bibinfo{author}{\bibfnamefont{R.}~\bibnamefont{Gandhi}},
  \bibinfo{author}{\bibfnamefont{A.}~\bibnamefont{Baishya}},
  \bibinfo{author}{\bibfnamefont{T.}~\bibnamefont{Santhosh}},
  \bibinfo{author}{\bibfnamefont{R.}~\bibnamefont{Tripathi}}, \bibnamefont{and}
  \bibinfo{author}{\bibfnamefont{T.}~\bibnamefont{Nag}},
  \bibinfo{journal}{Phys. Rev. C} \textbf{\bibinfo{volume}{104}},
  \bibinfo{pages}{L031602} (\bibinfo{year}{2021}).

\bibitem[{\citenamefont{Jurado and Schmidt}(2020)}]{schmidt16}
\bibinfo{author}{\bibfnamefont{B.}~\bibnamefont{Jurado}} \bibnamefont{and}
  \bibinfo{author}{\bibfnamefont{K.~H.} \bibnamefont{Schmidt}},
  \bibinfo{journal}{http://www.khs-erzhausen.de/GEF-2020-1-2.html Computer code
  {\sc gef}, Version 1.2}  (\bibinfo{year}{2020}).

\bibitem[{\citenamefont{Schmidt et~al.}(2016)\citenamefont{Schmidt, Jurado,
  Amouroux, and Schmitt}}]{kh16}
\bibinfo{author}{\bibfnamefont{K.-H.} \bibnamefont{Schmidt}},
  \bibinfo{author}{\bibfnamefont{B.}~\bibnamefont{Jurado}},
  \bibinfo{author}{\bibfnamefont{C.}~\bibnamefont{Amouroux}}, \bibnamefont{and}
  \bibinfo{author}{\bibfnamefont{C.}~\bibnamefont{Schmitt}},
  \bibinfo{journal}{Nucl. Data Sheets} \textbf{\bibinfo{volume}{131}},
  \bibinfo{pages}{107} (\bibinfo{year}{2016}).

\bibitem[{\citenamefont{S.Santra et~al.}(2022)\citenamefont{S.Santra, Pal, and
  et. al.}}]{ss22}
\bibinfo{author}{\bibnamefont{S.Santra}},
  \bibinfo{author}{\bibfnamefont{A.}~\bibnamefont{Pal}}, \bibnamefont{and}
  \bibinfo{author}{\bibnamefont{et. al.}}, \bibinfo{journal}{under review}
  (\bibinfo{year}{2022}).

\bibitem[{\citenamefont{Banerjee et~al.}(2022)\citenamefont{Banerjee, Kozulin,
  Burtebayev, Gikal, Knyazheva, Itkis, Novikov, Kvochkina, Mukhamejanov, and
  Pan}}]{tg22}
\bibinfo{author}{\bibfnamefont{T.}~\bibnamefont{Banerjee}},
  \bibinfo{author}{\bibfnamefont{E.~M.} \bibnamefont{Kozulin}},
  \bibinfo{author}{\bibfnamefont{N.~T.} \bibnamefont{Burtebayev}},
  \bibinfo{author}{\bibfnamefont{K.~B.} \bibnamefont{Gikal}},
  \bibinfo{author}{\bibfnamefont{G.~N.} \bibnamefont{Knyazheva}},
  \bibinfo{author}{\bibfnamefont{I.~M.} \bibnamefont{Itkis}},
  \bibinfo{author}{\bibfnamefont{K.~V.} \bibnamefont{Novikov}},
  \bibinfo{author}{\bibfnamefont{T.~N.} \bibnamefont{Kvochkina}},
  \bibinfo{author}{\bibfnamefont{Y.~S.} \bibnamefont{Mukhamejanov}},
  \bibnamefont{and} \bibinfo{author}{\bibfnamefont{A.~N.} \bibnamefont{Pan}},
  \bibinfo{journal}{Phys. Rev. C} \textbf{\bibinfo{volume}{105}},
  \bibinfo{pages}{044614} (\bibinfo{year}{2022}).

\bibitem[{\citenamefont{Mohanto et~al.}(2020)\citenamefont{Mohanto, De,
  Parihari, Rout, Ramachandran, Mahata, Mirgule, Gandhi, Sangeeta, Kushwaha
  et~al.}}]{gm20}
\bibinfo{author}{\bibfnamefont{G.}~\bibnamefont{Mohanto}},
  \bibinfo{author}{\bibfnamefont{S.}~\bibnamefont{De}},
  \bibinfo{author}{\bibfnamefont{A.}~\bibnamefont{Parihari}},
  \bibinfo{author}{\bibfnamefont{P.~C.} \bibnamefont{Rout}},
  \bibinfo{author}{\bibfnamefont{K.}~\bibnamefont{Ramachandran}},
  \bibinfo{author}{\bibfnamefont{K.}~\bibnamefont{Mahata}},
  \bibinfo{author}{\bibfnamefont{E.~T.} \bibnamefont{Mirgule}},
  \bibinfo{author}{\bibfnamefont{R.}~\bibnamefont{Gandhi}},
  \bibinfo{author}{\bibnamefont{Sangeeta}},
  \bibinfo{author}{\bibfnamefont{M.}~\bibnamefont{Kushwaha}},
  \bibnamefont{et~al.}, \bibinfo{journal}{Phys. Rev. C}
  \textbf{\bibinfo{volume}{102}}, \bibinfo{pages}{044610}
  (\bibinfo{year}{2020}).

\bibitem[{\citenamefont{K.F.Flynn et~al.}(1972)\citenamefont{K.F.Flynn,
  E.P.Horwitz, C.A.A.Bloomquist, R.F.Barnes, R.K.Sjoblom, P.R.Fields, and
  L.E.Glendenin}}]{flynn72}
\bibinfo{author}{\bibnamefont{K.F.Flynn}},
  \bibinfo{author}{\bibnamefont{E.P.Horwitz}},
  \bibinfo{author}{\bibnamefont{C.A.A.Bloomquist}},
  \bibinfo{author}{\bibnamefont{R.F.Barnes}},
  \bibinfo{author}{\bibnamefont{R.K.Sjoblom}},
  \bibinfo{author}{\bibnamefont{P.R.Fields}}, \bibnamefont{and}
  \bibinfo{author}{\bibnamefont{L.E.Glendenin}}, \bibinfo{journal}{Phys. Rev.
  C} \textbf{\bibinfo{volume}{5}}, \bibinfo{pages}{1725}
  (\bibinfo{year}{1972}).

\bibitem[{\citenamefont{Schmidt et~al.}(2000)\citenamefont{Schmidt,
  Steinhauser, Bockstiegel, Grewe, Heinz, Junghans, Benlliure, Clerc, de~Jong,
  Muller et~al.}}]{kh00}
\bibinfo{author}{\bibfnamefont{K.-H.} \bibnamefont{Schmidt}},
  \bibinfo{author}{\bibfnamefont{S.}~\bibnamefont{Steinhauser}},
  \bibinfo{author}{\bibfnamefont{C.}~\bibnamefont{Bockstiegel}},
  \bibinfo{author}{\bibfnamefont{A.}~\bibnamefont{Grewe}},
  \bibinfo{author}{\bibfnamefont{A.}~\bibnamefont{Heinz}},
  \bibinfo{author}{\bibfnamefont{A.}~\bibnamefont{Junghans}},
  \bibinfo{author}{\bibfnamefont{J.}~\bibnamefont{Benlliure}},
  \bibinfo{author}{\bibfnamefont{H.-G.} \bibnamefont{Clerc}},
  \bibinfo{author}{\bibfnamefont{M.}~\bibnamefont{de~Jong}},
  \bibinfo{author}{\bibfnamefont{J.}~\bibnamefont{Muller}},
  \bibnamefont{et~al.}, \bibinfo{journal}{Nuclear Physics A}
  \textbf{\bibinfo{volume}{665}}, \bibinfo{pages}{221} (\bibinfo{year}{2000}).

\bibitem[{\citenamefont{Mahata et~al.}(2021)\citenamefont{Mahata, Schmitt,
  Gupta, Shrivastava, Scamps, and Schmidt}}]{mahata22}
\bibinfo{author}{\bibfnamefont{K.}~\bibnamefont{Mahata}},
  \bibinfo{author}{\bibfnamefont{C.}~\bibnamefont{Schmitt}},
  \bibinfo{author}{\bibfnamefont{S.}~\bibnamefont{Gupta}},
  \bibinfo{author}{\bibfnamefont{A.}~\bibnamefont{Shrivastava}},
  \bibinfo{author}{\bibfnamefont{G.}~\bibnamefont{Scamps}}, \bibnamefont{and}
  \bibinfo{author}{\bibfnamefont{K.-H.} \bibnamefont{Schmidt}},
  \bibinfo{journal}{Phys. Lett. B} \textbf{\bibinfo{volume}{825}},
  \bibinfo{pages}{136859} (\bibinfo{year}{2021}).

\bibitem[{\citenamefont{Molnar and et. al.}(1990)}]{molner90}
\bibinfo{author}{\bibfnamefont{G.}~\bibnamefont{Molnar}} \bibnamefont{and}
  \bibinfo{author}{\bibnamefont{et. al.}}, \bibinfo{journal}{Symposium on the
  occasion of the 40th anniversary of the nuclear shell model}
  \textbf{\bibinfo{volume}{23}}, \bibinfo{pages}{23069047}
  (\bibinfo{year}{1990}).

\bibitem[{\citenamefont{Boboshin et~al.}(2007)\citenamefont{Boboshin, Varlamov,
  Ishkhanov, and et. al.}}]{bobo07}
\bibinfo{author}{\bibfnamefont{I.}~\bibnamefont{Boboshin}},
  \bibinfo{author}{\bibfnamefont{V.}~\bibnamefont{Varlamov}},
  \bibinfo{author}{\bibfnamefont{B.}~\bibnamefont{Ishkhanov}},
  \bibnamefont{and} \bibinfo{author}{\bibnamefont{et. al.}},
  \bibinfo{journal}{Phys. Atom. Nuclei} \textbf{\bibinfo{volume}{70}},
  \bibinfo{pages}{1363} (\bibinfo{year}{2007}).
\end{thebibliography}


\end{document}